# Algorithm of Amorphous Carbonaceous Nanomaterial Structure Identification with a Joint X-Ray and Neutron Diffraction Data Analysis

V.S. Neverov*, V.V. Voloshinov‡, A.B. Kukushkin*, A.S. Tarasov‡

* National Research Center "Kurchatov Institute", 123182, Moscow, Russia
‡ Institute for Information Transmission Problems of Russian Academy of Science, 127994, Moscow, Russia

Abstract. An algorithm is developed for structure identification of amorphous carbonaceous nanomaterials with a joint x-ray and neutron diffraction data analysis, using the data on the chemical composition of the sample from other diagnostics. The algorithm is an extension of one formerly developed for the x-ray diffraction diagnostics of similar nanomaterials [1]. The algorithm is tested on the example of processing the computed data for a sample which models a fragment of hydrocarbon films analyzed in [1]. The algorithm is of interest for structure diagnostics of materials with comparable atomic fractions of carbon and hydrogen.

## 1. INTRODUCTION

The structure identification of carbonaceous nanomaterials in the size range of 1-10 nm, using the X-ray diffraction (XRD) patterns in the range of scattering wave vector's modulus $q$ from several units to several tens of inverse nanometers, are of practical interest for characterization of nanometer-sized carbon nanoparticles in the powder carbon samples, films and various carbonaceous nanomaterials. Interpretation of the measurements of synchrotron XRD by the hydrocarbon films, deposited in the vacuum vessel of tokamak T-10, has requested massive calculations of the XRD patterns of multiple carbon nanoparticles of various topology and size [1]. These included the following carbon nanoparticles: fullerenes; carbon nanotubes (CNT) of various radius, chirality and length; axially symmetric ellipsoids (including the spheres); low-aspect-ratio toroids of elliptic cross-section; all possible half-fragments of all above-mentioned structures; various multilayered nanoparticles, including double- to five-wall flakes of the flat graphene, with the flake's position being decorrelated (i.e. glassy graphite). Implementation of the optimization procedure [1] became possible due to using the method [2] of an approximate description of the positions of carbon atoms in a curved graphene sheet. Method [2] is based on the local rearrangement of the neighboring atoms in the walls of hollow carbon sp2 structures with graphene-like wall, that enabled us to calculate the diffraction patters approximately without the knowledge of exact atomic positions in the wall. This method was implemented in the XaNSoNS (X-ray and Neutron Scattering on Nanoscale Structures) software package, available for remote use via RESTful web-services created in the Mathcloud (www.mathcloud.org) distributed environment. The RESTful web-services can be easily integrated into distributed computing scenarios. Such a scenario of X-ray diffraction data processing for carbonaceous nanomaterials has been presented in [3].

One of the peculiarities of structure identification problem for carbonaceous nanomaterials with a substantial amount of hydrogen, e.g. nanostructured hydrocarbon films, is the difficulty of calculation (and interpretation) of amorphous hydrocarbon contribution to diffraction patterns in the range of $q > 30$ nm$^{-1}$. The processing of neutron diffraction data for the same sample could be an important additional source of information. The joint structure refinement of x-ray and neutron diffraction data is widely applied for liquids and disordered condensed matter (see e.g. [4]). The feature of the problem we study here is the presence of an unknown molecular components (nanostructured carbon), which requires special identification methods [1] using the new method [2] for massive calculations of diffraction patterns for the above mentioned structures.

In the present paper we extend the algorithm of structure identification, formerly developed for the X-ray powder diffraction diagnostics of carbonaceous nanomaterials, to the case of a joint X-ray and neutron powder diffraction data processing, using the data on the chemical composition of the sample from other diagnostics (section 2). The algorithm is tested via processing the computed X-ray and neutron powder diffraction data for a sample which models a fragment of the hydrocarbon films analyzed in [1] (section 3).

## 2. AN ALGORITHM FOR STRUCTURE IDENTIFICATION WITH A JOINT X-RAY AND NEUTRON DIFFRACTION DATA ANALYSIS

The algorithm is defined as that for solving an optimization problem which is posed as a minimization of the deviation of the calculated scattering intensity from the measured one both for X-ray and neutron diffraction data. Here we consider the case of a homogeneous amorphous mixtures of purely carbon nanostructures, "free" carbon atoms (actually these atoms belong to a hydrocarbon amorphous component; in such a model we neglect the C-H correlations of positions of C and H atoms in hydrocarbon molecules), and impurities. Generalization of the algorithm presented below to a more complex chemical mixtures is straightforward. The quantities under minimization are as follows:

$$Z_j^{Xray}(\mathbf{x},a,b^{Xray},A) \equiv \frac{1}{S_{norm}^{Xray}} S_{\exp}^{Xray}(q_j) - \sum_{i=1}^{N} S_i^{Xray}(q_j) \cdot x_i - a \cdot S_{C\,amorph}^{Xray}(q_j) - A \cdot S_{impur\,amorph}^{Xray}(q_j) - b^{Xray}, \quad (j=1:m), \quad (1)$$

$$Z_k^{Neut}(\mathbf{y},c,b^{Neut},B) \equiv \frac{1}{S_{norm}^{Neut}} S_{\exp}^{Neut}(q_k) - \sum_{i=1}^{N} S_i^{Neut}(q_k) \cdot y_i - c \cdot S_{C\,amorph}^{Neut}(q_k) - B \cdot S_{impur\,amorph}^{Neut}(q_k) - b^{Neut}, \quad (k=1:n), \quad (2)$$

where the input data are as follows:
- $S_{\exp}^{Xray}(q_j)$ and $S_{\exp}^{Neut}(q_k)$ are the experimental X-ray and neutron scattering intensities, respectively;
- $S_i^{Xray}(q_j)$ and $S_i^{Neut}(q_k)$ are the calculated X-ray and neutron scattering intensities for the $i$-th carbon nanostructure, divided by the number of atoms in the structure;
- $j$ and $k$ are the current numbers of the points in the discrete spaces (generally different) of scattering vector modulus $q$ for X-ray and neutron data;
- $N$ is the total number of nanostructures selected for optimization;
- $S_{C\,amorph}^{Xray}(q_j)$ and $S_{C\,amorph}^{Neut}(q_k)$ are the calculated X-ray and neutron scattering intensities for amorphous carbon medium of given density, divided by total number of carbon atom in the sample;
- $S_{impur\,amorph}^{Xray}(q_j)$ and $S_{impur\,amorph}^{Neut}(q_k)$ are total contributions to the calculated X-ray and neutron scattering intensities from all the impurities (their contents in the sample is assumed to be known from the results of other diagnostics) for a given density of impurities in the sample, taken with the weight fractions equal to the ratios of impurity atoms number to the total number of carbon atoms in the sample;
- $m$ and $n$ are the total numbers of the points in the discrete spaces of scattering vector magnitude for X-ray and neutron data.

The output data for the optimization problem are as follows:
- $x_i$ and $y_i$ are the weight coefficients proportional to the number of carbon atoms in the purely carbon nanostructures for the calculated X-ray and neutron scattering intensities, respectively,
- $a$ and $c$ are the similar weight coefficients for "free" carbon atoms,
- $A$ and $B$ are the similar weight coefficients for all carbon atoms in the sample,

- $b^{Xray}$ and $b^{Neut}$ are the values of the possible constant background signal in the experimental X-ray and neutron scattering data.

As it follows from the above the ratios $x_i/A$ and $y_i/B$ are the probabilities of carbon atom to belong to the $i$-th nanostructure within optimal distributions of nanostructures predicted by the separate processing of the X-ray or neutron diffraction data, respectively; and the ratios $a/A$ и $c/B$ are the similar probabilities of carbon atom to belong to amorphous medium.

The values $S_{norm}^{Xray}$ and $S_{norm}^{Neut}$ are introduced to normalize the experimental data: $\sum_{j=1}^{m} S_{\exp}^{Xray}(q_j) \equiv S_{norm}^{Xray}$ and $\sum_{k=1}^{n} S_{\exp}^{Neut}(q_k) \equiv S_{norm}^{Neut}$. There are additional constraints on a constant background, which allow for the errors of experimental data:

$$\frac{1}{S_{norm}^{Xray}} S_{\exp}^{Xray}(q_j) - a \cdot S_{C\,amorph}^{Xray}(q_j) - A \cdot S_{impur\,amorph}^{Xray}(q_j) - b^{Xray} \geq -\frac{1}{S_{norm}^{Xray}} S_{\exp}^{Xray\,err} \quad (j=1:m), \qquad (3)$$

$$\frac{1}{S_{norm}^{Neut}} S_{\exp}^{Neut}(q_k) - c \cdot S_{C\,amorph}^{Neut}(q_k) - B \cdot S_{impur\,amorph}^{Neut}(q_k) - b^{Neut} \geq -\frac{1}{S_{norm}^{Neut}} S_{\exp}^{Neut\,err} \quad (k=1:n), \qquad (4)$$

where the values $S_{\exp}^{Xray\,err}$ и $S_{\exp}^{Neut\,err}$ describe the average errors of the experimental X-ray and neutron data. One may introduce more additional constraints related to the boundary values of these quantities:

$$b_{Low}^{Xray} \leq b^{Xray} \leq b_{Up}^{Xray},\ b_{Low}^{Neut} \leq b^{Neut} \leq b_{Up}^{Neut} \qquad (5)$$

The output variables obey the following conditions:

$$\sum_{i=1}^{N} x_i + a = A,\ \sum_{i=1}^{N} y_i + c = B,\ x_i, y_i \geq 0\ (i=1:N),\ a,c \geq 0. \qquad (6)$$

For the joint processing of the data for the same sample one has the following condition of compatibility:

$$\frac{x_i}{A} = \frac{y_i}{B}\ (i=1:N),\ \frac{a}{A} = \frac{c}{B}. \qquad (7)$$

Similarly to [1], we use three independent criteria of optimization in order to have an additional estimate of the accuracy of the final results:

$$K \sum_{j=1}^{m} \left| Z_j^{Xray}(\mathbf{x},a,b^{Xray},A) \right| + (1-K) \sum_{k=1}^{n} \left| Z_k^{Neut}(\mathbf{y},c,b^{Neut},B) \right| \xrightarrow[\mathbf{x,y},a,c,\mathbf{b},A,B]{} \min, \qquad \textbf{(L}_1\textbf{ case)}\quad (8)$$

$$K \sum_{j=1}^{m} \left( Z_j^{Xray}(\mathbf{x},a,b^{Xray},A) \right)^2 + (1-K) \sum_{j=1}^{m} \left( Z_k^{Neut}(\mathbf{y},c,b^{Neut},B) \right)^2 \xrightarrow[\mathbf{x,y},a,c,\mathbf{b},A,B]{} \min, \qquad \textbf{(L}_2\textbf{)} \quad (9)$$

$$K \max_{j=1:m} \left| Z_j^{Xray}(\mathbf{x},a,b^{Xray},A) \right| + (1-K) \max_{k=1:n} \left| Z_k^{Neut}(\mathbf{y},c,b^{Neut},B) \right| \xrightarrow[\mathbf{x,y},a,c,\mathbf{b},A,B]{} \min. \qquad \textbf{(L}_{inf}\textbf{)} \quad (10)$$

Here $0 \leq K \leq 1$, the weight coefficient introduced by the user to allow for relative confidence of diagnostics which provide experimental data.

Equation (7) makes the optimization problem of Eqs. (1)-(10) a complicated nonlinear problem. We will return back to a problem of the same complexity as in [1] if take into account that values *A* and *B*, which describe integral quantity of carbon in the sample, are less sensitive to the results of joint processing of X-ray and neutron data as compared to the distribution over topology and size of nanostructures (i.e. over variable *i*). Under this assumption the algorithm will take the following form.

On the first step one has to carry out the minimization for each diagnostic separately, without Eq. (7). For all three criteria the respective optimization will be as follows:

$$\sum_{j=1}^{m} \left| Z_j^{Xray}(\mathbf{x}, a, b^{Xray}, A) \right| \xrightarrow[\mathbf{x}, a, b^{Xray}, A]{} \min \, , \quad (L_1) \quad (11)$$

$$\sum_{k=1}^{n} \left| Z_k^{Neut}(\mathbf{y}, c, b^{Neut}, B) \right| \xrightarrow[\mathbf{y}, c, b^{Neut}, B]{} \min \, , \quad (L_1) \quad (12)$$

$$\sum_{j=1}^{m} \left( Z_j^{Xray}(\mathbf{x}, a, b^{Xray}, A) \right)^2 \xrightarrow[\mathbf{x}, a, b^{Xray}, A]{} \min \, , \quad (L_2) \quad (13)$$

$$\sum_{k=1}^{n} \left( Z_k^{Neut}(\mathbf{y}, c, b^{Neut}, B) \right)^2 \xrightarrow[\mathbf{y}, c, b^{Neut}, B]{} \min \, , \quad (L_2) \quad (14)$$

$$\max_{j=1:m} \left| Z_j^{Xray}(\mathbf{x}, a, b^{Xray}, A) \right| \xrightarrow[\mathbf{x}, a, b^{Xray}, A]{} \min \, , \quad (L_{inf}) \quad (15)$$

$$\max_{k=1:n} \left| Z_k^{Neut}(\mathbf{y}, c, b^{Neut}, B) \right| \xrightarrow[\mathbf{y}, c, b^{Neut}, B]{} \min \, . \quad (L_{inf}) \quad (16)$$

Each criterion gives its own values of **x**, **y**, *a*, *c*, *A* and *B*. Let us fix the respective values $A_{L1}$, $A_{L2}$, $A_{Linf}$, and $B_{L1}$, $B_{L2}$, $B_{Linf}$, where the subscripts denote the criterion. On the second step these values, namely the pairs ($A_{L1}$, $B_{L1}$), ($A_{L2}$, $B_{L2}$) and ($A_{Linf}$, $B_{Linf}$), are used in solving the optimization problems of Eqs. (8), (9), and (10), respectively, with Eq. (7) taking the form of a linear one. The results of these optimizations include the effect of the joint data processing the data.

We use the LPSOLVE linear optimization package for $L_1$ and $L_{inf}$ criteria (Eq. (8), (10)-(12), (15), (16)) and the IPOPT non-linear solver for criterion $L_2$.

## 3. VERIFICATION OF THE ALGORITHM ON A MODEL HYPOTHETICAL SAMPLE

In order to test the algorithm of Sec. 2., we numerically simulated a "phantom" experimental X-ray and neutron diffraction patterns for a model sample composed from nanostructures in an amorphous medium with structural composition obtained by solving an inverse problem of X-ray diffraction data processing for hydrocarbon films from vacuum vessel of tokamak T-10 [1]. This model sample is shown in Fig. 1, and its structural content is given in Table 1 (see the "Assumed" column).

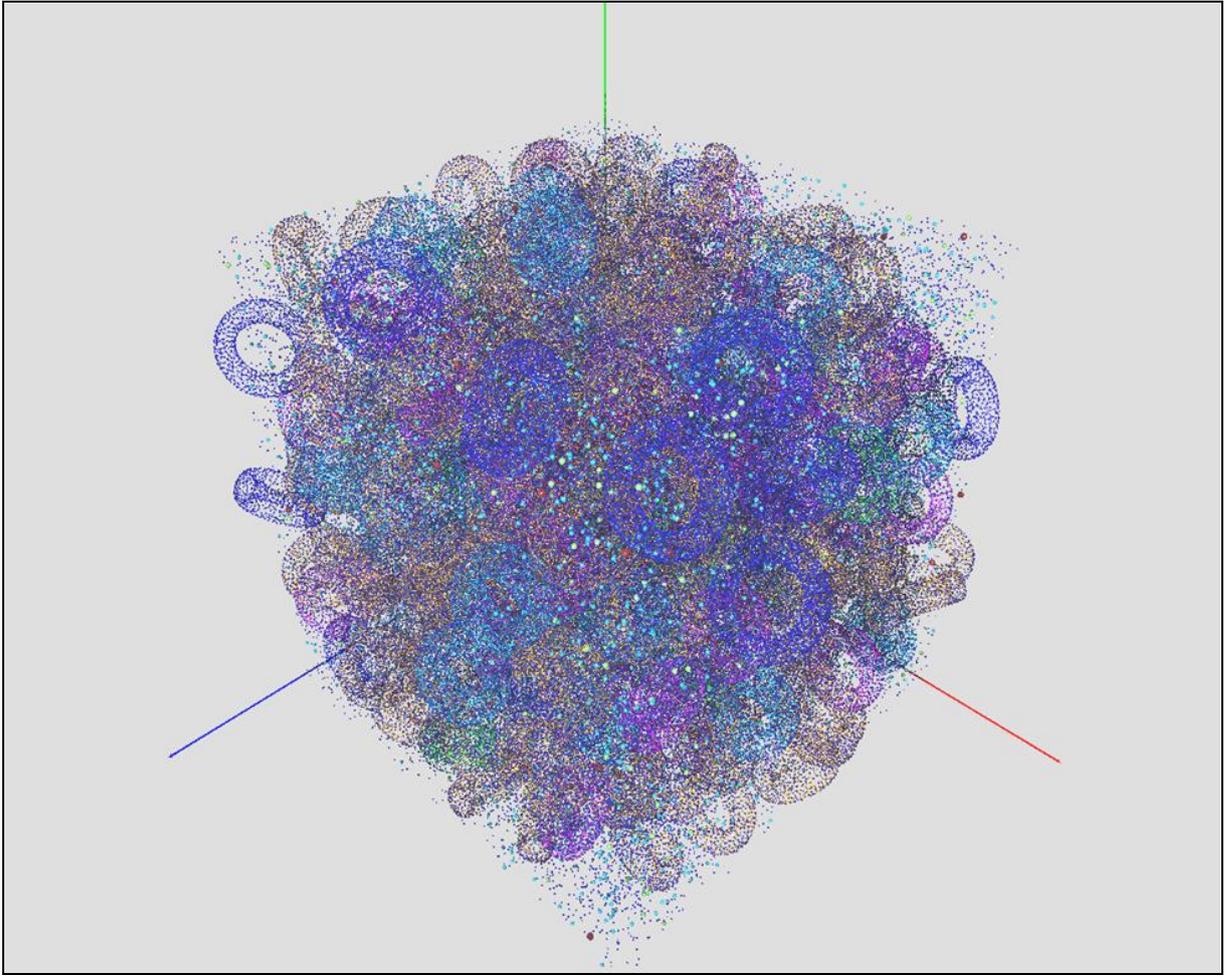

Figure 1. A phantom sample which models a fragment of hydrocarbon films, with small amount of metal impurities, analyzed in [1] (see Table 1 for structural content). Hydrogen and deuterium atoms are not shown.

We add a random error to the calculated X-ray and neutron scattering intensities in the following way:

$$S_{\exp}^{Xray}(q_j) \to S_{\exp}^{Xray}(q_j) + \max_{j=1:m}\left(S_{\exp}^{Xray}(q_j)\right) \cdot rand(-0.05, 0.05), \quad j = 1:m. \tag{17}$$

$$S_{\exp}^{Neut}(q_k) \to S_{\exp}^{Neut}(q_k) + \max_{k=1:n}\left(S_{\exp}^{Neut}(q_k)\right) \cdot rand(-0.05, 0.05), \quad k = 1:n. \tag{18}$$

We use the same set of 575 candidate nanostructures as we used for processing of hydrocarbon films in [1]. This set includes the following single-wall carbon structures:
i. fullerene $C_{60}$;
ii. irregular clusters of fullerenes, $C_{60\_n}$ (n=10, 20, …, 100, 120, 240);
iii. carbon nanotubes with variable radius (0.36 nm <R< 0.96 nm), respective allowable chirality, and length (L > 2$R_{min}$ ~ 7 nm);
iv. axially symmetric ellipsoids (including the spheres), 0.5 nm ≤ $R_x$ ≤ 0.7 nm, $R_x$ ≤ $R_y$ ≤ 2 $R_x$, where $R_x$ and $R_y$ are the minor and major radii;
v. toroids of elliptic cross section, 0.85 nm ≤ R ≤ 1.25 nm, 0.3 nm ≤ $R_x$ ≤ 0.4 nm, 0.5 nm ≤ $R_y$ ≤ 0.9 nm, where R is the major radius of toroid;
vi. all possible half-fragments of all the structures of items (iii) – (v).

Also, the following multi-layered nanostructures are included:

vii. from double- to four-wall carbon nanotubes, similarly to item (iii) and their half-fragments;
viii. from double- to five-wall flakes of the flat graphene of the size from 32 to 200 atoms per layer, with the flake`s position decorrelated (i.e. glassy graphite).

The results of the joint X-ray and neutron data processing for $L_2$ optimization criteria (9) and weight coefficient $K=0.5$ are shown in Fig. 2 and Fig. 3. The results of minimization for other criteria are slightly worse.

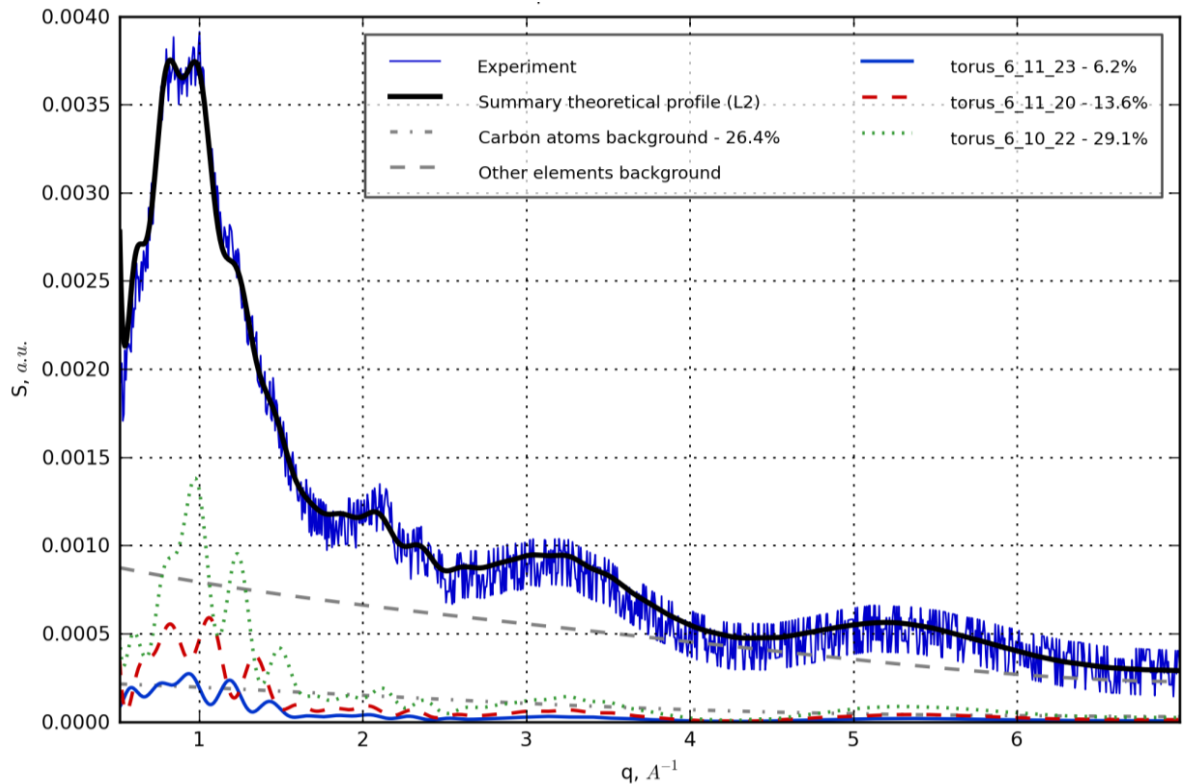

Figure 2. Comparison of theoretical X-ray intensity, obtained by minimizing the sum of squared deviations ($L_2$ criteria) with the simulated, "phantom" experimental data for weight coefficient $K=0.5$ (joint X-ray + neutron data processing). The most significant partial contributions to theoretical curve are indicated (see Table 1, last column).

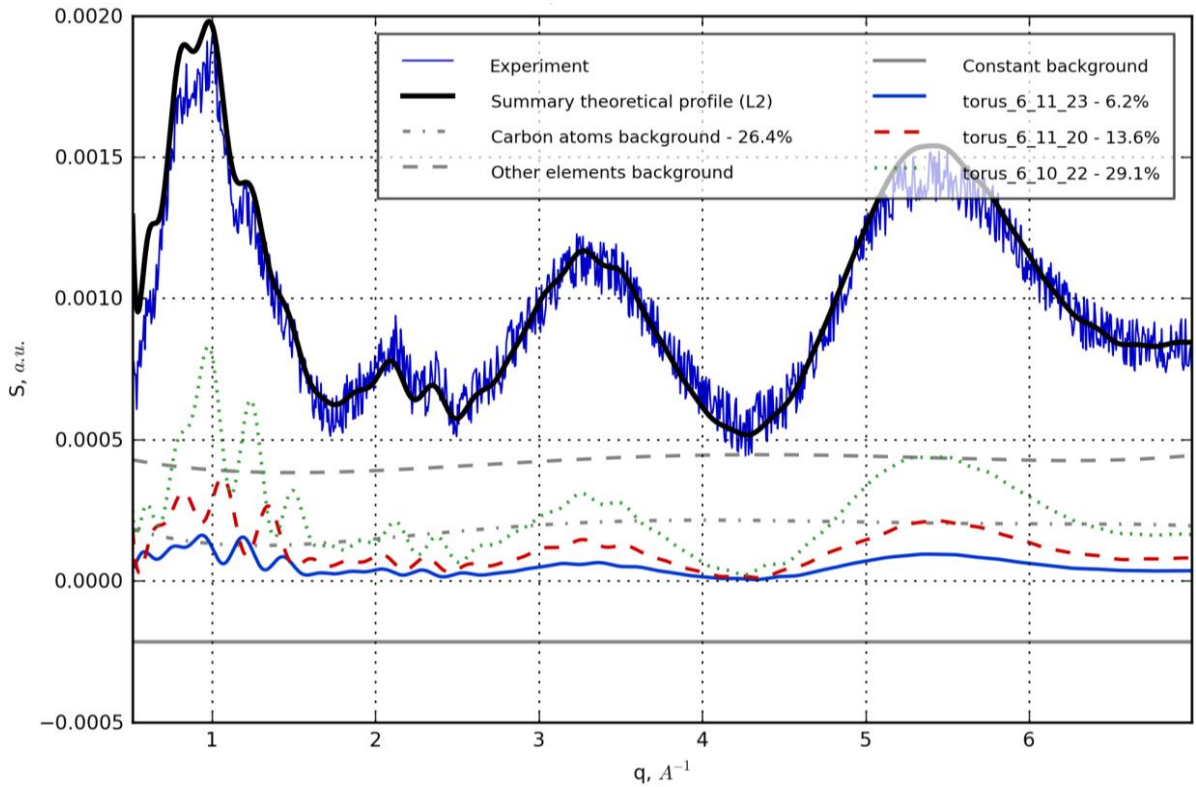

Figure 3. The same as in Fig. 2, but for neutron scattering.

The assumed and recovered structural contents of the model sample for $L_2$ optimization criteria are shown in Table 1 for different values of $K$ weight coefficient: X-ray data processing only ($K=1$), neutron data processing only ($K=0$), and the joint X-ray and neutron data processing ($K=0.5$). There is some difference of results for different values of $K$, however on the whole the influence of the joint optimization on the topological content is not strong.

One can see that the accuracy of the joint minimization is higher than that for neutron analysis only, but lower than for the X-ray analysis only. This suggests that the calculation error of the diffraction patterns for the amorphous hydrocarbon medium has a larger effect on the neutron diffraction patterns of the sample than on that for X-rays.

Table 1. Comparison of the results of solving the optimization problem for different values of the weight coefficient $K$ for optimization criteria $L_2$.

| Structure | | Assumed, % (>1%) | Recovered, % (>1%) | | |
|---|---|---|---|---|---|
| | | | $K=1$ (X-ray only) | $K=0$ (neutron only) | $K=0.5$ (joint analysis) |
| Disordered cluster of 240 $C_{60}$ molecules | | 5.6 | 2.5 | 1.8 | - |
| Toroidal CNT | $R_x = 3$ Å, $R_y = 5$ Å, $R = 9.5$ Å | - | 1.0 | - | 1.8 |
| | $R_x = 3$ Å, $R_y = 5$ Å, $R = 10$ Å | 12.9 | 12.6 | 12.5 | 3.1 |
| | $R_x = 3$ Å, $R_y = 5$ Å, $R = 10.5$ Å | 4.4 | - | 2.9 | 2.4 |
| | $R_x = 3$ Å, $R_y = 5$ Å, $R = 11$ Å | 36.6 | 36.4 | 27.2 | 29.1 |
| | $R_x = 3$ Å, $R_y = 5$ Å, $R = 12$ Å | 7.4 | 3.2 | 6.0 | 1.5 |
| | $R_x = 3$ Å, $R_y = 5.5$ Å, $R = 9.5$ Å | - | - | - | 1.35 |
| | $R_x = 3$ Å, $R_y = 5.5$ Å, $R = 10$ Å | - | 4.0 | - | 13.6 |
| | $R_x = 3$ Å, $R_y = 5.5$ Å, $R = 11.5$ Å | - | 3.3 | 2.6 | 6.2 |
| | $R_x = 3$ Å, $R_y = 6$ Å, $R = 12$ Å | - | 4.0 | - | - |
| | $R_x = 3$ Å, $R_y = 6.5$ Å, $R = 9.5$ Å | 13.2 | 4.2 | 2.1 | 2.2 |
| | $R_x = 3$ Å, $R_y = 6.5$ Å, $R = 12$ Å | - | | - | - |
| | $R_x = 3.5$ Å, $R_y = 6$ Å, $R = 9.5$ Å | - | 2.2 | - | - |
| | $R_x = 3.5$ Å, $R_y = 6$ Å, $R = 10$ Å | - | 2.9 | 1.2 | - |
| CNT, $R=4.92$ Å, chirality angle= 16.1° | | - | - | 1.1 | - |
| "Free" carbon | | 19 | 20.9 | 34.6 | 26.4 |

## 4. CONCLUSIONS

Here we developed an algorithm for structure identification of amorphous carbonaceous nanomaterials with a joint x-ray and neutron diffraction data analysis. The algorithm assumes using the data on the chemical composition of the sample from other diagnostics. The algorithm extends that formerly developed for the x-ray diffraction diagnostics of similar nanomaterials [1]. The algorithm is implemented for a wide class of carbon nanostructures. The algorithm is tested on the example of processing the computed data for a sample which models a fragment of hydrocarbon films analyzed in [1]. The algorithm is of interest for structure diagnostics of materials with a comparable atomic fractions of carbon and hydrogen.

As far as the software implementation of the joint X-ray and neutron diffraction data processing is completely analogous to the previously developed in [3] for the X-ray diffraction data processing, this new algorithm can be easily implemented as a distributed system of the RESTful-services (operating in accordance with the architectural style of REST) via the MathCloud software toolkit.

## ACKNOWLEDGEMENTS

The authors are grateful to P.V. Minashin, V.A. Rantsev-Kartinov, and P.A. Sdvizhenskii for the assistance, V.G. Stankevich and N.Yu. Svechnikov, for helpful discussion of experimental data for hydrocarbon films. V.A. Somenkov and A.A. Skovoroda, for stimulating our interest in the neutron diffraction diagnostics. The authors are also grateful to A.P. Afanasiev for his support of collaboration between the NRC "Kurchatov Institute" and the Center of Grid-Technologies and Distributed Computing (http://dcs.isa.ru), and Sukhoroslov O.V., for the support of computing infrastructure used in our research.

This work is supported by the Russian Foundation for Basic Research (project RFBR № 12-07-00529-a).